\documentclass[pra,amsmath,amssymb,showpacs,twocolumn]{revtex4}

 \usepackage{graphicx}
\usepackage{epsfig}
 \usepackage[latin1]{inputenc}
 \usepackage{xspace}
 \usepackage{latexsym}
 \usepackage{amsmath}
 \usepackage{amssymb}
 \usepackage{amstext}
 \usepackage{amsxtra}
 \usepackage{makeidx}

\def\poscal#1#2{\left\langle#1,#2\right\rangle}
\def\R{\mathbb{R}}
\def\Z{\mathbb{Z}}

\def\ep{\varepsilon}

\begin{document}

\title{Fast rotating condensates in an asymmetric harmonic trap}
\author{Amandine Aftalion}
\affiliation{CNRS-UMR7641, CMAP, Ecole Polytechnique, 91128
Palaiseu, France}
\author{Xavier Blanc}
\affiliation{Universit{\'e} Pierre et Marie Curie-Paris6, UMR
  7598, laboratoire Jacques-Louis Lions, 175 rue du Chevaleret, Paris
  F-75013 France}
\author{Nicolas Lerner}
\affiliation{Universit{\'e} Pierre et Marie Curie-Paris6, UMR
  7586, Institut de Mathématiques de Jussieu, 175 rue du Chevaleret, Paris
  F-75013 France}
\date{\today}

\begin{abstract} We investigate the effect of the anisotropy of a
harmonic trap on the behaviour of a fast rotating Bose-Einstein
condensate. Fast rotation is reached when the rotational velocity is
close to the smallest trapping frequency, thereby deconfining the
condensate in the corresponding direction. We characterize a regime
of velocity
  and small anisotropy  where
  the behaviour is similar to the isotropic case: a triangular Abrikosov
 lattice of vortices, with an inverted parabola profile.
 Nevertheless, at sufficiently large velocity, we find that the ground state does
 not display vortices in the bulk.
    We show that the coarse grained
atomic density behaves like an inverted parabola with large radius
in the deconfined direction, and keeps a fixed profile given by a
Gaussian in the other direction. The description is made within the
lowest Landau level set of states, but using distorted complex
coordinates.

\end{abstract}

\maketitle

Vortices appear in many quantum systems such as superconductors and
superfluid liquid helium. Rotating atomic gaseous Bose-Einstein
condensates constitute a novel many body system where vortices have
been observed \cite{Madison00} and various aspects of macroscopic
quantum physics can be studied. In a harmonically trapped condensate
rotating at a frequency close to the trap frequency, interesting
features have emerged, presenting a strong analogy with quantum Hall
physics. In the mean field regime, vortices form a triangular
Abrikosov lattice \cite{Abr} and the coarse grained density
approaches an inverted parabola \cite{CKR,WBP,ABD}. At very fast
rotation, when the number of vortices becomes close to the number of
atoms, the states are strongly correlated and the vortex lattice is
expected to melt \cite{Cooper01}. In the mean field regime, Ho
\cite{H} observed that the low lying states in a symmetric 2D trap
are analogous to those in the lowest Landau level (LLL) for a
charged particle in a uniform magnetic field. This analogy allows a
simplified description of the gas by the location of  vortices: the
wave function describing the condensate  is a Gaussian multiplied by
an analytic function  of the complex variable $z= x+ i y$. The
zeroes of the analytic function are the location of the vortices. It
is the distortion of the vortex lattice  on the edges of the
condensate which allows to create an inverted parabola profile
\cite{CKR,WBP,ABD,ABN2} for the coarse grained atomic density in the
LLL.

The experimental achievement of rotating BEC involves anisotropic
traps. An anisotropy of the trap can drastically change the picture
 in the fast rotation regime. In this case,
 the condensate becomes
 very elongated in one direction and forms a novel quantum fluid in
 a narrow channel. The investigation of the vortex pattern has been
 performed for an infinite strip which corresponds to the situation
 where the rotational frequency has reached the smallest trapping
 frequency \cite{gora,palacios}, and  for an elongated condensate
 \cite{LNF,oktel,fetter07}.
As pointed out by Fetter \cite{fetter07}, the description of the
condensate can still be made in the framework of the lowest Landau
level, defined by an anisotropic Gaussian, multiplied by an analytic
function of $x+i\beta y$, where $\beta$ is related to the anisotropy
of the trap and the rotational frequency.
We are going to characterize a regime of fast rotation where there
are no vortices in the bulk of the condensate and show that the
coarse grained density profile is very different from the isotropic
case: the behaviour is an inverted parabola with large expansion in
the deconfined direction, while the extension remains fixed
 in the other direction, with a
 Gaussian profile.

 We consider a 2D gas of $N$  atoms  rotating at frequency $\Omega$  around the
$z$ axis. The gas is confined in a harmonic potential, with
frequencies $\omega_x =\omega \sqrt{1-\nu^2}$, $\omega_y =\omega
\sqrt{1+\nu^2}$ along the $x,y$ axis respectively. The state of the
gas is described by a macroscopic wave function $\psi$ normalized to
unity, which minimizes the Gross-Pitaevskii energy functional. In
the following, we choose $\omega$, $\hbar \omega$, and
$\sqrt{\hbar/(m\omega)}$, as units of frequency, energy and length,
respectively. The dimensionless coefficient $G =  N a_s /a_z$
characterizes the strength of atomic interactions (here $a_s$ is the
atom scattering length and $a_z$ the extension of the wave function
in the $z$ direction for the initial 3-dimensional problem). The
energy in the rotating frame is
\begin{equation} E[\psi]=\int \left( \psi^*
\left[H_\Omega\psi\right] + \frac{G}{2} |\psi|^4  \right) \; dxdy
 \label{energy1}
 \end{equation}
where $H_\Omega$ is defined by
 \begin{eqnarray}
H_\Omega &=&-\frac{1}{2} \nabla^2 + \frac{1-\nu^2}{2}x^2 +
\frac{1+\nu^2}2 y^2 -\Omega
L_z 
 \label{singlepartH}
 \end{eqnarray}
and $L_z=i(y\partial_x-x\partial_y)$ is the angular momentum.  We
are going to study the fast rotation regime where $\Omega^2$
approaches the critical velocity $\Omega_c^2:=1-\nu^2$ from below.
Thus, we define the small parameter $\ep$ by $ \ep^2 = 1-\nu^2
-\Omega^2.$
 The spectrum of the Hamiltonian \eqref{singlepartH} has a Landau
level structure. The lowest Landau level is defined as (see \cite{fetter07})
\begin{equation}
  \label{eq:LLL3}
  f(x+i\beta y) e^{\left[-\frac \gamma {8\beta}\left(x^2 + (\beta
      y)^2\right) \right]
-i\frac{\nu^2}{2\Omega}xy}, \ f \text{ is
  analytic}
\end{equation} where $\gamma$ and $\beta$ are some constants
related to $\Omega$ and $\nu$ given in the appendix; $\beta$ is
close to 1 if $\nu$ is small.
 For such functions, $<{H_\Omega \psi},{\psi}>$ can be
 simplified (see the appendix and \cite{fetter07}), and in the small $\ep$ limit (with
 $\ep\ll \nu$), we are left with the study of
\begin{equation}
E_{LLL}(\psi)=   \int \frac12 \left(
  \ep^2 x^2 + \kappa^2 y^2\right) |\psi|^2+\frac G2 |\psi|^4 dx dy
  \label{eq:nrjLLL2}
\end{equation} where $\kappa^2\sim
(\nu^2+\ep^2/2)(2-\nu^2)/(1-\nu^2).$ This energy only depends on the
modulus of $\psi$. Hence, it is possible to forget the phase of
$\psi$, and use a simplified  definition of the LLL:
\begin{multline}
  \label{eq:LLL4}
  \psi(x,y) = f(x+i\beta y)
  e^{\left[-\frac{\gamma}{8\beta}\left(x^2+\beta^2 y^2\right) \right]},
  \ f\text{ is analytic.}
\end{multline}
We recall that the orthogonal projection of $L^2(\R^2)$ onto the LLL
is explicit \cite{B}:
 $ \Pi_{LLL} (\psi) = \frac \gamma {4\pi} \int
  e^{{-\frac{\gamma}{8\beta}\left(|z|^2 -2 z\overline{z'} +
 |z'|^2\right)}} \psi(x',y')dx'dy',$
where $z=x+i\beta y$ and $z'=x'+i\beta y'$. We refer to the appendix
of \cite{R} for details on the operator $\Pi_{LLL}$, its kernel and
the computations: if an LLL function $\psi$ (i.e $\psi$ satisfies
\eqref{eq:LLL4}) is the ground state of
 \eqref{eq:nrjLLL2}, it is a solution of the projected
Gross-Pitaevskii equation:
\begin{equation}
  \label{eq:GPLLL}
  \Pi_{LLL}\left[\left(\frac{\ep^2}2 x^2 + \frac{\kappa^2}2 y^2 +
      G|\psi|^2  -\mu\right)\psi \right] = 0,
\end{equation}
where $\mu$ is the chemical potential.

The ground state of (\ref{eq:nrjLLL2})  without the analytic
 constraint is the inverted parabola
\begin{equation}\label{TF}
|\psi|^2 = \rho_{\rm TF}: = \frac{2}{\pi R_x R_y}\left(1 -
\frac{x^2}{R_x^2} -
  \frac{y^2}{R_y^2} \right),
\end{equation}
where
$R_x =\left(\frac{4G \kappa}{\pi\ep^3}\right)^{1/4}$, $R_y =
\left(\frac{4G \ep}{\pi \kappa^3}\right)^{1/4}.$
Note that in the isotropic case $\nu=0$ (that is $\kappa = \ep$), one
recovers the standard circular
  shape $R_x = R_y =[4G/(\pi\ep^2)]^{1/4}$.
  Since $\kappa\gg \ep$, $R_x$ is always large. On the other hand, the behaviour of $R_y$
  depends on the respective values of $\ep$ and $\kappa\sim \nu \sqrt2$.
  We find that $R_y$ is large if
   $\nu \ll \ep^{1/3}$ while $R_y$ shrinks if $\nu \gg \ep^{1/3}$.
   We are going to see that in the first case, the
profile \eqref{TF} is reached in the fast rotation limit in the LLL
using a vortex lattice, exactly as in the isotropic case, while in
the second case, \eqref{TF} is not a good description of the
condensate because the properties of the LLL prevent $R_y$  from
shrinking, and in particular the energy is much higher than that of
\eqref{TF}.

\begin{figure}[htb]
 \begin{centering}{\includegraphics[width=4cm]{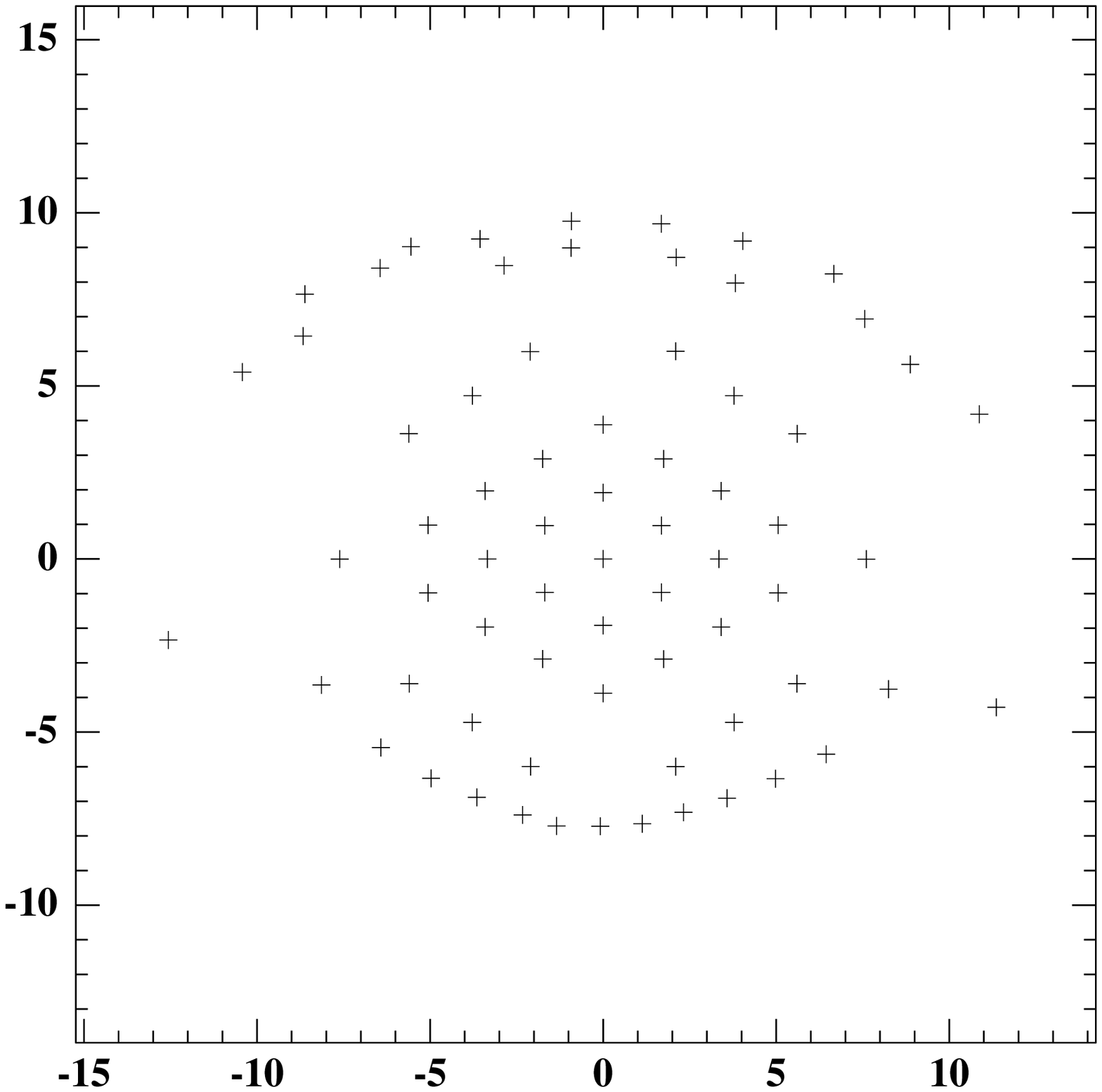}
\includegraphics[width=4.5cm]{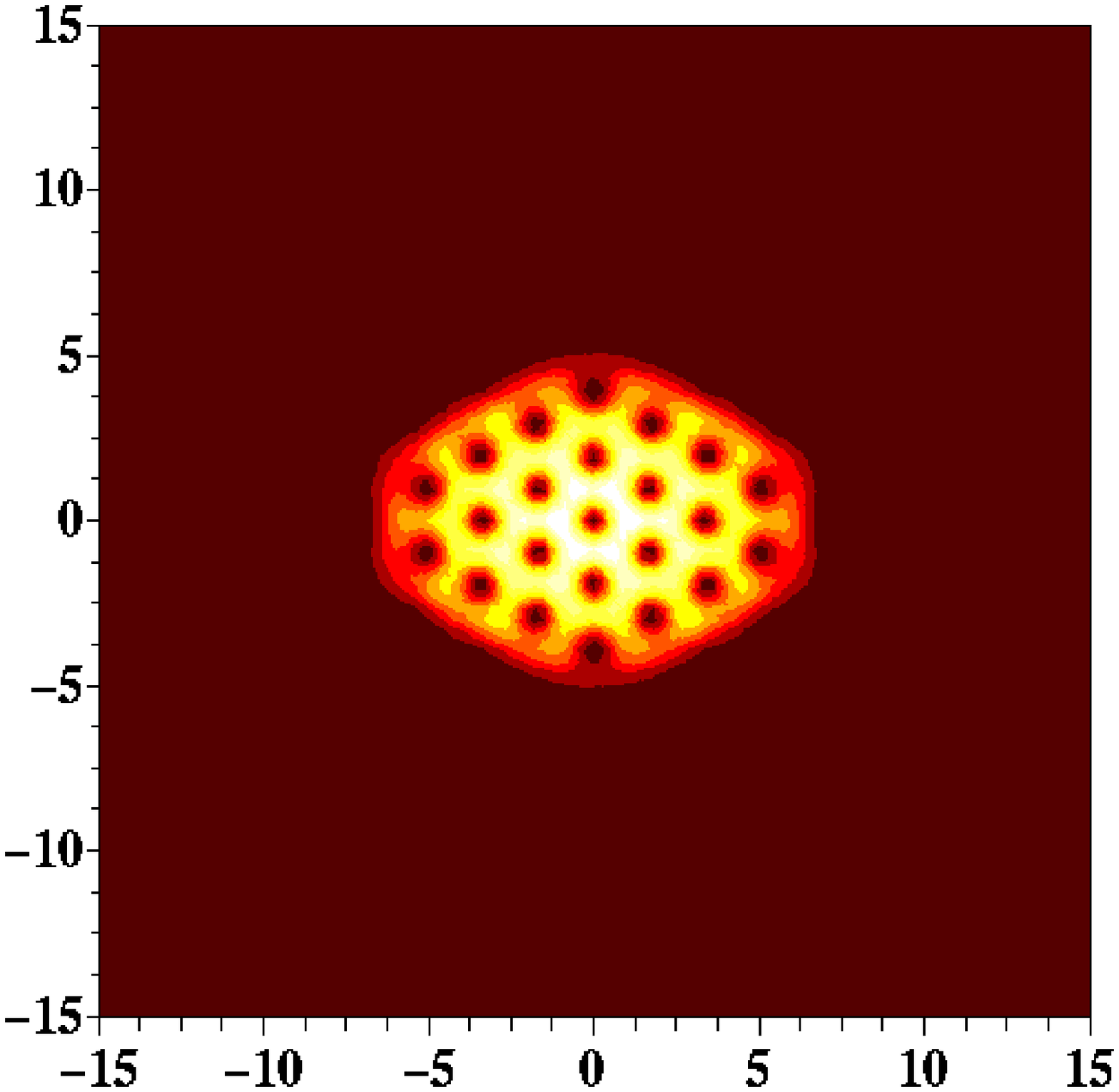}}\end{centering}
 \caption{\label{fig2}An example of  (a): a configuration of the zeroes
 (b): density plot. There are 58 vortices with 23 visible vortices.
  $\nu=0.03,$ $\Omega = 0.9985$, $\ep^2=2\times 10^{-3}$, $G=3$.}
\end{figure}

In the first regime $\nu \ll \ep^{1/3}$, which we  call the weakly
anisotropic case, figure~\ref{fig2} provides a typical vortex
configuration, together with the corresponding density plot. It is
obtained by minimizing the energy as a function of the location of
vortices $z_i$ with a conjugate gradient method.
 The vortex lattice can be described as in the Abrikosov problem
  \cite{Abr,H}
using the Theta function:
\begin{equation}\label{theta}
\phi(x,y ; \tau) = e^{\frac{\gamma}{8\beta}\left({z^2} -
{|z|^2}\right)} \Theta \left(\sqrt{\frac{\tau_I\gamma}{4\pi\beta
      }}z, \tau \right),
\end{equation}
where $z = x + i\beta y$ and $\tau=\tau_R+i\tau_I$ is the lattice
parameter. The zeroes of the function $\phi $ lie on the lattice
$\sqrt{\frac{4\pi\beta}{\tau_I\gamma}}\left(\Z\oplus \Z\tau\right)$
and $|\phi|$ is periodic. The optimal lattice, that is the one
minimizing $b(\tau)=\int |\phi|^4/(\int |\phi|^2)^2$
 is triangular, which corresponds to $\tau =
e^{2i\pi/3}$ (the integrals are taken on one period).
 As in the isotropic case \cite{ABN2}, we can
construct an approximate ground state of \eqref{eq:nrjLLL2} by
multiplying the solution \eqref{theta} of the Abrikosov problem by a
profile $\rho$ varying at the same scale as $\rho_{\rm TF}$  defined
in \eqref{TF}.
Since this product is not in the LLL, we project it onto the LLL and
define $v = \Pi_{LLL} \left( \rho(x,y) \phi(x,y;\tau)\right)$ whose
energy is
$$
E_{\rm LLL}(v) =  \int_{\R^2} \left(\frac{\ep^2}2
 x^2 + \frac{\kappa^2}2 y^2\right) \rho
+\frac {Gb(\tau)} 2  \rho^2dxdy,
$$
up to an error of order $ \sqrt{\kappa\ep}
  (\kappa^3/\ep)^{1/8}.$
  Then, minimizing with respect to $\rho$
 yields that
$\rho(x,y) = \frac 1 {\sqrt{b(\tau)}} \rho_{\rm TF}
  \left(\frac{x}{b(\tau)^{1/4}}, \frac{y}{b(\tau)^{1/4}}\right)$
  where $\rho_{\rm TF}$ is given by \eqref{TF}.
The condensate indeed expands in both directions, and a
coarse-grained density profile is close to the anisotropic inverted
parabola.
 The vortex lattice is not distorted by the
 anisotropy since $\beta$ is close to 1; it
is still triangular, as displayed in figure~\ref{fig2}.
Nevertheless, as in the isotropic case \cite{ABD,ABN2}, the lattice
is distorted on the edges of the condensate, thereby allowing for a
coarse-grained Thomas-Fermi profile in the LLL description.

\begin{figure}[t]
 \begin{centering}{\includegraphics[width=4cm]{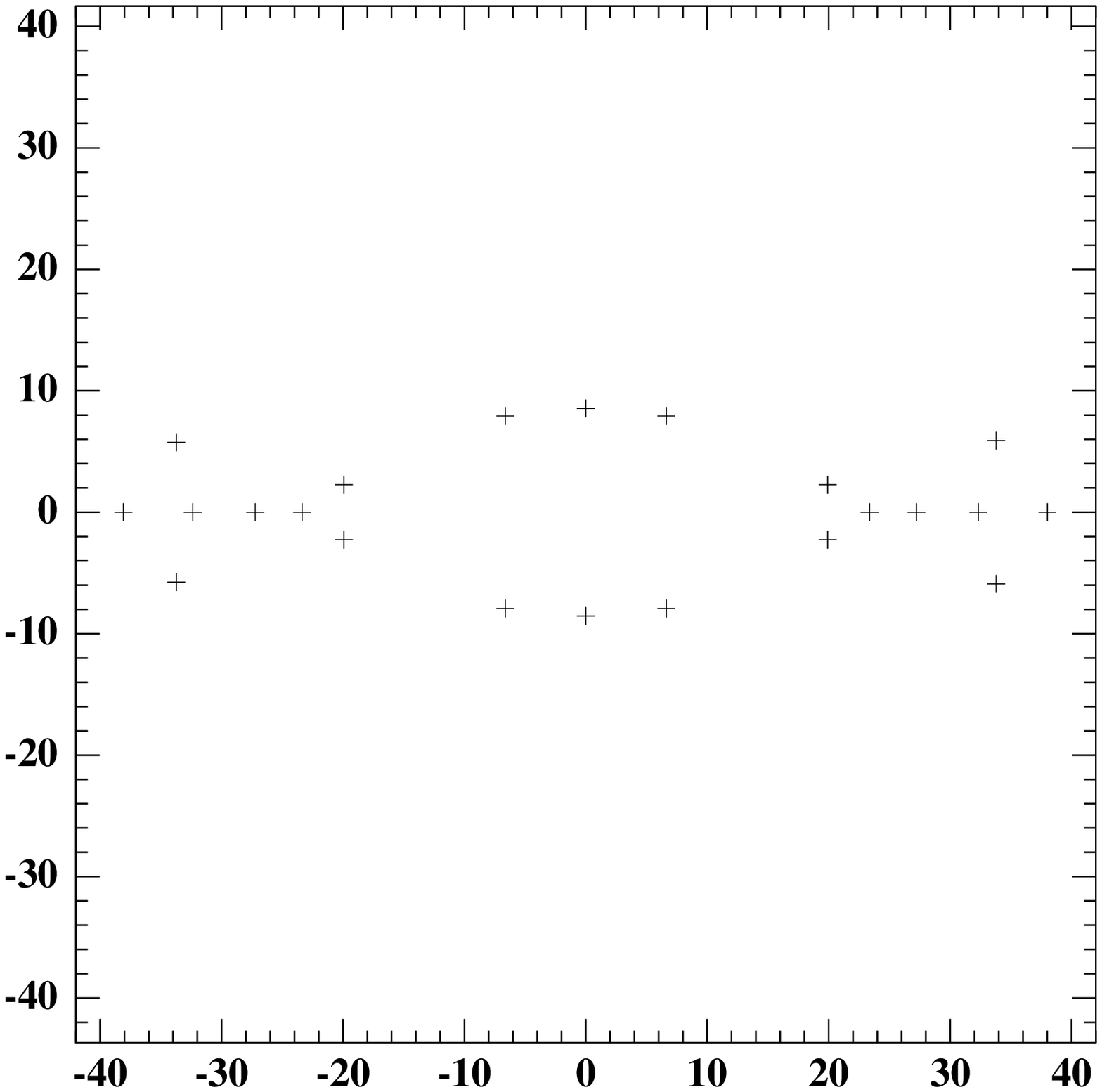}
\includegraphics[width=4.5cm]{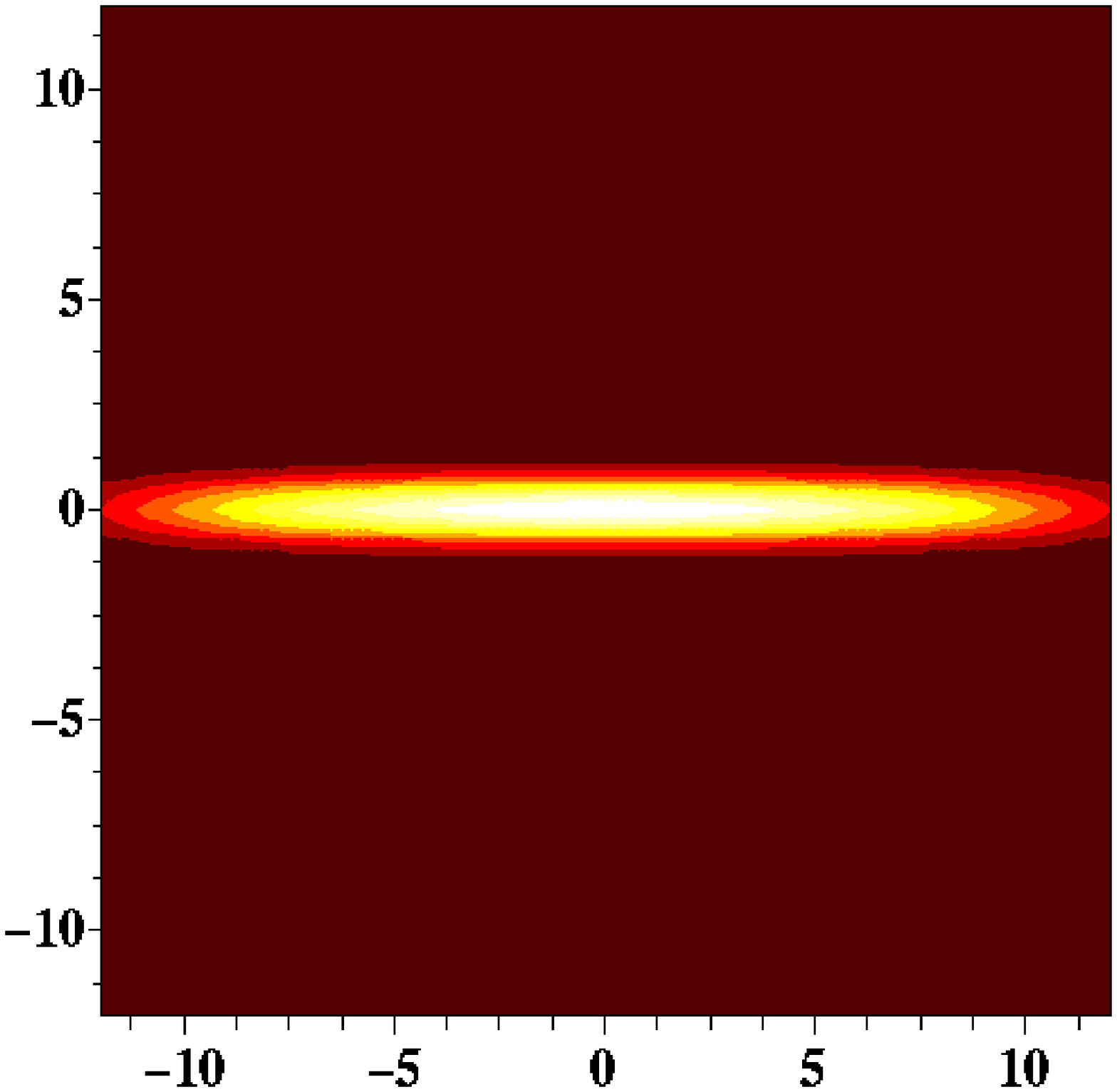}}\end{centering}
 \caption{\label{fig:fort}An example of  (a): a configuration of the zeroes
  (b): density
   plot. There are only invisible vortices (32 vortices).
   Here, $\nu=0.73$ $\Omega = 0.6820$, $\ep^2=2\times 10^{-3}$, $G=3$. The
   extension in the $y$ direction is given by (\ref{eq:lim})}
\end{figure}

 However, when $\nu \gg \ep^{1/3}$, that is for fast rotation,
 the behaviour is very different as illustrated in Figure \ref{fig:fort}:
 we are going to see that the ground state
    is close to a Gaussian in the $y$ direction
multiplied by an inverted parabola in the $x$ direction. There is no
vortex lattice. There are only invisible vortices whose role is to
create the profile in the LLL. The function
  (\ref{TF}) does not provide the correct behaviour of the ground state: though $R_y$ in
  (\ref{TF}) is small, the condensate does not shrink in the $y$
  direction
   but keeps a fixed Gaussian profile\cite{foot2}
\begin{equation}\label{eq:lim} g(x,y) =\left (
{\frac{\gamma\beta}{2\pi}}\right )^{1/4}
\exp\left(-\frac{\gamma\beta}{4} y^2 +
  i\frac{\gamma}{4} xy\right).
\end{equation}
We are going to prove that if $u$ is the ground state and $p(x)$ its
projection onto the Gaussian \eqref{eq:lim},
 then $p(x)$ is almost an inverted parabola.
 Indeed, in the LLL, we have the key identity $\int (\partial_x
|\psi|)^2+ (1/\beta^2) (\partial_y |\psi|)^2=(\gamma/4\beta)\int
|\psi|^2$ (see \cite{carlen}). By adding and subtracting
$\kappa^2/(2\gamma\beta)\int|\psi|^2$ to the energy, and using this
identity, we find
$$E_{LLL}(\psi)=-\frac{\kappa^2}{2\gamma\beta}+\int \left (
\frac{2\kappa^2}{\gamma^2\beta^2} (\partial_y
|\psi|)^2+\frac{\kappa^2}2 y^2 |\psi|^2\right )$$ $$+\int \left (
\frac{2\kappa^2}{\gamma^2} (\partial_x |\psi|)^2+\frac{\ep^2}2 x^2
|\psi|^2+\frac G2 |\psi|^4\right ).$$ This expression of the energy
allows to analyze separately the contributions in the $x$ and $y$
directions. The ground state of
$-(2/\gamma^2\beta^2)\partial_{y}^2+(1/2)y^2$ is the modulus of
(\ref{eq:lim}) and the ground energy is $1/(\gamma\beta)$.
 Projecting any function of the LLL onto the space generated
 by (\ref{eq:lim}) times a function of $x$, and using that
 $\ep^{1/3}/\nu$ is small, we find that $E_{LLL}(u)\geq
 (\kappa^2/(2\gamma\beta))+E_{1D}(p(x))$ where
 \begin{equation}\label{ep1D}E_{1D}(p)=\int_\R  \left(\frac{2\kappa^2}{\gamma^2}{(p')}^2+\frac 12
  \ep^2 x^2 p^2  + \frac G 4 \sqrt{\frac{\gamma
      \beta}{\pi}} p^4 \right)dx.\end{equation}
       The minimizer of
      $E_{1D}$ among all $p$'s is of Thomas-Fermi type and we call it $q$:
\begin{equation}\label{TF2}
q(x) = \sqrt{\frac{3}{4R}}\left(1 - \frac{x^2}{R^2} \right)^{1/2}_+
\ , \quad R = \left(\frac{3G}{4\ep^2}\sqrt{\frac{\gamma \beta}{\pi}}
\right)^{1/3}
\end{equation} since $\ep^{1/3}/\nu$ is
      small.
       This gives the energy estimate
\begin{equation}
  \label{eq:bsup}
\min  E_{LLL} -  \frac{\kappa^2}{ 2 \gamma \beta } \geq
E_{1D}(q)\sim \frac3{10}\left(\frac{3\ep}4 G
    \sqrt{\frac{\gamma\beta}{\pi}} \right)^{2/3}.
\end{equation}

 Let us point out that this lower bound is optimal since we can
 construct a test function in the LLL with this energy.  We project  a Dirac delta function  in the $y$
direction times an inverted parabola in $x$, that is $v(x,y) =
A\Pi_{LLL} [ \delta_0(y) q(x)]$, where $q$ is the function
\eqref{TF2}:
\begin{equation}
  \label{eq:fct}
  v(x,y)
= \frac{A\gamma}{4\pi} e^{-\frac{\gamma\beta}{8} y^2 } \int_{\R} e^{
  -\frac{\gamma}{8\beta } \left((x-x')^2 - 2ix'\beta y \right)}
 q(x')dx'.
\end{equation}
  The constant $A =
(2\pi/\gamma\beta)^{1/4}$ is a normalization factor. The fact that
$q$ varies on a scale of order $\ep^{-2/3}$ allows to expand
\eqref{eq:fct} in powers of $\ep^{2/3}$:
\begin{multline}\label{DL}
v(x,y) =\left(\frac{\gamma \beta}{2\pi}\right)^{1/4} q(x)
\exp\left(-\frac{\gamma\beta}{4}y^2 +
  i\frac\gamma{4}xy\right)  \\
+ \ep^{2/3}\left(\frac{\gamma \beta}{2\pi}\right)^{1/4} q'(x) iy
\exp\left(-\frac{\gamma\beta}{4}y^2 +
  i\frac\gamma{4}xy\right)
\end{multline} with an error of order
$\ep^{4/3}$.
 Inserting this expansion in the energy, we find
\begin{equation}
  E_{\rm LLL}(v) =\frac{\kappa^2}{ 2 \gamma \beta }+\int_\R  \left(\frac 12
  \ep^2 x^2 q(x)^2  + \frac G 4 \sqrt{\frac{\gamma
      \beta}{\pi}} q(x)^4 \right)dx
\end{equation} with an error of order $\ep^{4/3}$.
This matches our lower bound \eqref{eq:bsup}. Let us point out that
according to \eqref{DL}, the wave function $v$ has no  vortices in
the bulk. This is corroborated by the numerical computation
displayed in Figure~\ref{fig:fort}. Nevertheless, the inverted
parabola profile in the $x$ direction is obtained in the LLL thanks
to the existence of invisible vortices, that is vortices outside the
support of this parabola.

Let us point out that the operator $y^2$, whose ground state is the
Gaussian \eqref{eq:lim} is bounded below by a positive constant in
the LLL: $\int_{\R^2} y^2 |\psi(x,y)|^2 dxdy \geq
 \frac{1}{\gamma \beta} \int_{\R^2} |\psi|^2$. This can be viewed as a kind of uncertainty principle
 \cite{bound}. This decoupling in the $x$ and $y$ directions  is possible only when the leading
 order term in the energy
 $\kappa^2/(\gamma\beta)\sim\nu^2$,  is larger than
 the energy of \eqref{TF} $\sqrt{G\nu \ep}$, that is when the
 ratio $\nu^3/\ep$ is large. When $\nu^3/\ep$ becomes
 of order 1, all the terms in the energy seem
  of the same order, the decoupling in the $x$ and $y$ variables is
  no longer meaningful, and \eqref{TF} does not provide the good
  behaviour either. The analysis in this intermediate regime is still
  open:
 it could display rows of vortices
   as obtained by \cite{palacios}.

The estimate of the energy \eqref{eq:bsup} allows us to justify the
 validity of the model: indeed,
 the mean field
approximation is valid if the number $N$ of particles is much larger
than the number of one-particle states allowed by the chemical
potential $\mu$, that is $N \gg  \mu/\mu_1$. Thanks to
\eqref{eq:bsup} and \eqref{eq:nrj-lll2}, we find $N \gg G^{2/3} /
\epsilon^{1/3} \nu$. Since $G$ is of order 1, $\epsilon^2 \sim
10^{-3}$ and $\nu \leq 10^{-1}$, this criterion is satisfied as long
as $N$ is greater than $10^4$, which corresponds to actual values in
experiments.  However, if $\ep$ gets too small, this condition gets
violated and the states get correlated. The LLL approximation is
valid if the 1D energy $E_{1D}$ is much smaller than the gap $\mu_2$
between the LLL and the first excited state: $(G \ep)^{2/3} \ll 1$.

{\em Conclusion:}  When the anisotropy is small compared to how
close the rotational velocity is to the critical velocity, that is
$\ep^{1/3} \gg \nu$, the behaviour is similar to the isotropic case
with  a triangular vortex lattice. A striking new feature is the
non-existence of visible vortices for the ground state of the energy
in the fast rotation regime, that is when $\ep^{1/3}\ll \nu$. The
profile of the ground state is a large inverted parabola in the
deconfined direction and a fixed Gaussian in the other direction.
   Our analysis indicates that an asymmetric rotating condensate
   undergoes a similar transition as a condensate placed in a
   quadratic+quartic trap where at large rotation the bulk of the
   condensate does not display vortices\cite{JFS}.
  Our investigation  opens new prospects for the
experiments: in particular, if a condensate at rest is set to
sufficiently large rotation, then vortices should not be nucleated.




\section*{Appendix} As computed in \cite{fetter07} on the basis of
ideas of Valatin \cite{valatin},
 the
eigenvalues of the Hamiltonian $H_\Omega$ are $1,\mu_1^2,1,
\mu_2^2$, where $ \mu_1^2 = 1+\Omega^2 - \sqrt{\nu^4 + 4\Omega^2},$
$ \mu_2^2 = 1+ \Omega^2 + \sqrt{\nu^4 + 4\Omega^2}.$ We define
$\alpha= \sqrt{\nu^4 + 4\Omega^2},$ $\beta_{1}=({2\Omega
\mu_{1}})/({\alpha-2\Omega^2+\nu^2}),$ $\beta = \beta_{2}=({2\Omega
\mu_{2}})/({\alpha+2\Omega^2+\nu^2}),$
$\gamma=({2\alpha})/{\Omega},$ $\lambda_{1}^2=
({\alpha-2\Omega^2+\nu^2})/({2\alpha}),$
$\lambda_{2}^2=({\alpha+2\Omega^2+\nu^2})/{2\alpha},$ $d=({\gamma
\lambda_{1}\lambda_{2}})/{2},$
$c=({\lambda_{1}^2+\lambda_{2}^2})/{2\lambda_{1}\lambda_{2}}.$ Then
$H_\Omega = \frac12\left(a_1^\dagger a_1 + a_1a_1^\dagger \right) +
\frac12\left(a_2^\dagger  a_2 + a_2a_2^\dagger \right)$ where $ a_2
= \frac{\mu_2}{\sqrt 2}\left(-i\lambda_1d^{-1}
\partial_{x} + c \lambda_1 y\right ) + \frac i{\sqrt 2}
\left(-i\lambda_2 \partial_{y} - \left(d\lambda_1^{-1} - \lambda_2
    cd \right)x \right)$, and
$a_1 = \frac{\mu_1}{\sqrt 2}\left(-i\lambda_2d^{-1}\partial_{y} +
c\lambda_2x \right) + \frac i{\sqrt 2} \left((\lambda_1 cd -
d\lambda_2^{-1})y -i \lambda_1 \partial_{x} \right).$ We have:
$\left[a_2, a_2^\dagger \right] = {\mu_2}, \quad \left[a_1,
a_1^\dagger \right] = {\mu_1},$ and all other commutators vanish.
The LLL is defined by  $a_2\psi =0,$ that is $
  f(x+i\beta_2 y) e^{\left[-\frac 1 {8\beta_2}\left(\frac{2\alpha -
        \nu^2}\Omega x^2 + \frac{2\alpha + \nu^2}\Omega (\beta_2
      y)^2\right) \right]
-i\frac{\nu^2}{4\Omega}xy}$, with $f$
  analytic.
It is always possible to change the analytic function $f(\xi)$ into
$f(\xi) \exp(-\delta\xi^2)$ in the above definition, since
$\exp(-\delta\xi^2)$ is an analytic function of $\xi$. Hence, for
$\delta = \nu^2/(8\Omega\beta_2),$ we find the alternative
definition of the LLL (\ref{eq:LLL3}), with $\beta=\beta_2$. This definition is equivalent
to the one given by Fetter in \cite{fetter07}. However, contrary to
\cite{fetter07}, the coefficients in \eqref{eq:LLL3} are not
singular in the limit $\ep\to 0.$ Indeed, in this limit, $\beta_2
\sim \sqrt{(1-\nu^2)/(1-\nu^2/2)}$ and $\gamma \sim
(4-2\nu^2)/\sqrt{1-\nu^2}$. This is due to the addition of the
above-mentionned complex Gaussian in the definition of the LLL.
In the LLL, we have
$\poscal{H_\Omega \psi }{\psi} =\frac12 \poscal{\left(a_1^\dagger
a_1 + a_1
  a_1^\dagger \right)\psi}{\psi} + \frac{\mu_2}2
  \poscal{\psi}{\psi}.$
We then express $x$ and $y$ as linear combinations of
$a_1,a_2,a_1^\dagger,a_2^\dagger$ \cite{fetter07,oktel} and get, if
$\psi\in LLL$,
\begin{multline}\label{eq:nrj-lll2}
\poscal{H_\Omega \psi}{\psi} = \frac{\mu_2}2 -
\frac{\mu_1}{4}\left(\beta_1\beta_2 +
  \frac1{\beta_1\beta_2} \right)  \\
+\frac \gamma 4  \int \left(
  \mu_1\beta_1 x^2 + \frac{\mu_1}{\beta_1} y^2\right) |\psi|^2 dx dy
\end{multline}
which provides (\ref{eq:nrjLLL2}) with $\kappa^2=\gamma
\mu_1/2\beta_1$ since $\gamma\mu_1\beta_1\sim 2\ep^2$. Note that
$\mu_1\sim \nu\ep$ and $\mu_2\sim (2-\nu^2)$.

\par\noindent
{\bf Acknowledgements.} We would like to thank A.L.Fetter and
J.Dalibard for very useful comments. We also acknowledge support
from the French ministry grant {\it ANR-BLAN-0238, VoLQuan} and
express our gratitude to our colleagues  participating to this
 project, in particular Th.Jolic\oe ur and S.Ouvry.


\begin{thebibliography}{99}
\bibitem{foot2}{This Gaussian is both
   a solution to $\Pi_{LLL} ( y^2 \psi) = c_1 \psi$ and $\Pi_{LLL} ( |\psi|^2
   \psi)= c_2 \psi$ for some constants $c_1,c_2$.}
\bibitem{bound} If $\psi\in LLL$, and $|\psi (x,y)|\leq A
e^{-ax^2}e^{-by^2}$, then our uncertainty principle implies a bound
on $b$, which means that an $LLL$ function cannot shrink in the $y$
direction.
\end{thebibliography}
\end{document}